\pdfoutput=1
\documentclass[usenatbib]{mn2e}
\usepackage{apjfonts}
\usepackage{amssymb}
\usepackage{amsmath}
\usepackage{ctable}
\usepackage{fixltx2e} 

\newcommand{\be}{\begin{equation}}
\newcommand{\ee}{\end{equation}}

\newcommand{\msun}{M_{\sun}}

\newcommand{\fgas}{f_{\rm gas}}

\newcommand{\scaleup}{}

\newcommand\plotone[1]
 {\centering \leavevmode \includegraphics[width={0.99\columnwidth}]{#1}}

\newcommand\altaffilmark[1]{$^{#1}$}
\newcommand\altaffiltext[1]{$^{#1}$}
\title[Dynamical AGN-Starburst Delays]{
\vspace{-0.25in}
Dynamical Delays Between 
Starburst and AGN Activity in Galaxy Nuclei
\vspace{-0.25in}
}

\author[Hopkins]{
\parbox[t]{\textwidth}{ 
Philip F.~Hopkins\thanks{E-mail:phopkins@astro.berkeley.edu}\altaffilmark{1} 
}
\vspace*{6pt} \\
\altaffiltext{1}{Department of Astronomy and Theoretical Astrophysics Center, University of California Berkeley, Berkeley, CA 94720} 
\vspace{-0.25in}
}

\date{\vspace{-0.25in}
Submitted to MNRAS, January, 2011}
\begin{document}
\maketitle
\label{firstpage}

\begin{abstract}
Observations of AGN have suggested a possible delay between the peak of 
star formation (on some scale) and AGN activity. Inefficient fueling (and/or feedback) 
from fast stellar winds has been invoked to explain this, but 
we argue this is unlikely in 
bright systems accreting primarily cold dense gas. 
We show that such a delay can arise even in bright quasars for purely dynamical 
reasons. If some large-scale process produces rapid inflow, smaller scales will quickly 
become gas-dominated. As the gas density peaks, so does the SFR. 
However, gravitational torques which govern further inflow are relatively inefficient in 
gas-dominated systems; as more gas is turned into stars, the stars provide an 
efficient angular momentum sink allowing more rapid inflow. 
Moreover, the gas provided to the central regions in mergers or strong 
disk instabilities will typically be $\sim100$ times larger than that needed to fuel the BH; 
the system is effectively in the ``infinite gas supply'' limit. 
BH growth can therefore continue for some time while the gas supply exhausts, 
until it has significantly depleted to the point where the BH is finally ``starved.'' 
Both of these effects act together with comparable magnitude, and 
mean that the peak of BH growth can lag the peak in the SFR 
measured at a given scale by a timescale corresponding to the gas exhaustion time 
on that scale ($\sim 10-100$ local dynamical times). 
This predicts that the inferred delay will vary in a specific manner with the 
radius over which the star formation rate is measured. 
We discuss possible implications for the role of 
AGN feedback in suppressing star formation activity. 
\end{abstract}

\begin{keywords}
galaxies: formation --- galaxies: evolution --- galaxies: active --- 
quasars: general --- cosmology: theory
\vspace{-0.25in}
\end{keywords}

\vspace{-0.7cm}
\section{Introduction}
\label{sec:intro}

The idea of a connection between AGN and starburst activity in 
galaxies has a long history, going back to speculation that 
dense nuclear gas concentrations in 
ultra-luminous infrared galaxies (ULIRGs) fuel super-massive 
black holes that eventually appear as luminous quasars, expelling 
the nuclear gas and dust \citep[][]{sanders:agn.vs.sf.in.ulirgs,
joseph:sb.vs.agn.power.ulirgs}. 
The discovery of tight correlations between the mass of nuclear black holes 
and their host bulge properties \citep{magorrian,
FM00,Gebhardt00,hopkins:bhfp.obs,
aller:mbh.esph,feoli:bhfp.1} 
has suggested that black hole and galaxy formation are co-eval, 
and the cosmic history of black hole growth 
has roughly similar shape to the cosmic 
star formation history \citep[see e.g.][]{merloni:magorrian.evolution}. 
Observations have found AGN to 
preferentially live in hosts with enhanced 
star formation activity relative to control galaxy samples \citep{brotherton99:postsb.qso,
canalizostockton01:postsb.qso.mergers,
kauffmann:qso.hosts,jahnke:qso.host.sf,
sanchez:qso.host.colors,vandenberk:qso.spectral.decomposition,
kaviraj:2011.agn.fb.needed.in.postsb,
zakamska:qso.hosts}. 

However, some recent observations have suggested that AGN and 
starburst activity may not be contemporaneous, even in a 
statistical sense. 
Specifically, 
\citet{davies:sfr.properties.in.torus} study a local sample of AGN at high 
resolution and see tentative suggestions that strong AGN are not present 
until the stellar populations in the central $\lesssim10-100\,$pc are of order 
a few $10^{7}\,$yr old. On large scales, \citet{schawinski:agn.sf.suppression.timescale} 
and \citet{wild:agn.sf.offset} \citep[see also][]{trichas:2009.sb.agn.lir.vs.lx,
tadhunter:2011.sb.agn.delay} compare the 
level of AGN activity to the integrated (galaxy-wide 
or central-kpc, respectively) stellar 
populations of local galaxies, and suggest that 
AGN activity may increase after the global stellar populations reach an age of a 
few $10^{8}$\,yr.

\citet{norman:stellar.wind.fueling} suggested that a delay between SF and AGN 
activity might result from stellar evolution 
timescales if the BH is fueled directly by stellar winds.
At low black hole accretion rates (BHARs), this may be important: young stellar populations 
($\lesssim100\,$Myr) 
produce fast but tenuous O-star 
winds, which contribute to a diffuse ISM that is hot and/or 
has large turbulent/bulk motions of $\gg100\,{\rm km\,s^{-1}}$. 
If the BHAR is dominated by the accretion of gas from 
this hot/diffuse phase of the ISM, then the Bondi-Hoyle accretion rate 
is suppressed by a factor $\propto c_{\rm eff}^{-3}$, where 
$c_{\rm eff}$ is the effective sound speed and/or turbulent speed in the ISM. 
As a consequence, if this energy cannot be efficiently radiated, 
larger accretion rates will require the arrival of 
a higher fraction of ``slow winds'' from more evolved stellar populations, 
which (having lower velocities and higher densities) can be more easily accreted.
A stronger extension to this argument might even be that 
the energy/momentum feedback in these winds suppresses accretion even of colder gas for some additional 
time \citep{wild:agn.sf.offset}.
There are suggestions that accretion at very low Eddington 
ratios, $\lesssim10^{-3}\,L_{\rm Edd}$, may be dominated 
by accretion of diffuse/hot gas, in which case the nature of 
stellar ejecta could have dramatic effects \citep{allen:jet.bondi.power,
soria:adaf.candidates.stellarwinds,hickox:multiwavelength.agn}. 

However, at high accretion rates $\gtrsim 1\%$ of Eddington, 
it is generally believed that BHs are accreting from a 
dynamically {\em cold, disky} high-density medium.
For example, maser observations have shown 
typical material at $\sim1-10\,$pc has densities $\gtrsim10^{8}\,{\rm cm^{-3}}$, 
with effective temperatures $\lesssim 100\,$K and 
turbulent velocities of $\sim10-50\,{\rm km\,s^{-1}}$ 
\citep{greenhill:circinus.acc.disk,lonsdale:vlbi.agn.cores,fruscione:abs.by.warped.disk,
henkel:agn.masers,kondratko:agn.masers.2,kondratko:ngc3393.acc.disk.maser,
nenkova:clumpy.torus.model.2}. 
The cold, disky, dense phase exists in e.g.\ dense starburst systems 
on $\sim100-1000\,$pc scales regardless of the presence of extremely young 
stellar populations \citep{downes.solomon:ulirgs,bryant.scoville:ulirgs.co,
tacconi:ngc6240.gasdynamics}. 
This is not surprising:  tenuous winds will 
have little dynamical/accelerative effect on cold clouds at these densities, 
and the gas is believed to be driven in from larger radii, not {locally} produced 
by stellar ejecta.
For typical conditions, with $M_{\ast}\sim M_{\rm gas}$ on scales $\lesssim 10\,$pc, 
the mass/momentum injection rates from O-star winds are much
smaller than the inflow rate/gravitational force 
(for a nuclear SFR similar to observed values in \citealt{davies:sfr.properties.in.torus} and 
wind momentum-loading from \citealt{starburst99}).
Moreover, even if the cold medium is replenished or heated by winds, the cooling time in such regions is 
$\sim 10^{-5}$ times shorter than the dynamical time, so the medium can immediately 
re-form a cold disk \citep[it retains little memory of the wind properties; see e.g.][]{schartmann:2010.1068.star.cluster.fueling}.

In this paper, we therefore investigate whether ``delays'' similar to those observed can 
appear independent of stellar fueling (or feedback) processes. We show, using 
a series of high-resolution hydrodynamic 
simulations, that basic dynamical processes can give rise to qualitatively 
similar delays even in AGN accreting from cold, dense media. 



\vspace{-0.7cm}
\section{The Simulations}
\label{sec:sims}

\citet{hopkins:zoom.sims} give a detailed description of the
simulations used here; we briefly summarize some 
important properties.  The simulations were performed with the
parallel TreeSPH code {\small GADGET-3} \citep{springel:gadget}.
They include stellar disks and bulges, dark
matter halos, gas, and BHs.  

Because of the large dynamic range in both space and time needed
for the self-consistent simulation of galactic inflows and nuclear
disk formation, we use a ``zoom-in'' re-simulation approach.  This
begins with a large suite of simulations of galaxy-galaxy mergers, and
isolated bar-(un)stable disks.  These simulations have
$0.5\times10^{6}$ particles, corresponding to a spatial resolution of
$50\,$pc. These simulations have been described in a series of previous papers 
\citep{dimatteo:msigma,robertson:msigma.evolution,
  cox:kinematics,younger:minor.mergers,hopkins:disk.survival}.

Following gas down to the BH accretion disk requires much higher
spatial resolution than is present in the galaxy-scale simulations. We
therefore select snapshots from the galaxy-scale simulations at key
epochs and 
isolate the central $0.1-1$\,kpc region, which contains
most of the gas that has been driven in from large scales.\footnote{Typically
$\sim10^{10}\,\msun$ of gas, within 
a scale length of $\sim0.3-0.5\,$kpc.}
From this mass distribution, we then re-populate the gas in the
central regions at much higher resolution, and simulate the dynamics
for many local dynamical times. Specifically, we either 
take the mass distribution in e.g.\ the central kpc ``as is'' from the 
parent distribution (``cutting out'' the 
central region and re-populating it with higher-resolution particles), 
or fit it to an exponential disk model which we initialize as an idealized BH/disk/bulge system in equilibrium 
(to avoid artificial features from the low-resolution initial simulation). 
These simulations involve $10^{6}$
particles, with a resolution of a few pc and particle masses of
$\approx 10^{4}\,\msun$.  We have run $\sim50$ such re-simulations,
corresponding to variations in the global system properties, the model
of star formation and feedback, and the exact time in the larger-scale
dynamics at which the re-simulation occurs.
\citet{hopkins:zoom.sims} present a number of tests of this
re-simulation approach and show that it is reasonably robust for this
problem.  This is largely because, for gas-rich disky systems, the
central $\sim 300$ pc becomes strongly self-gravitating, generating
instabilities that dominate the subsequent dynamics.


These initial re-simulations capture the dynamics down to $\sim 10$
pc, still insufficient to quantitatively describe accretion onto a
central BH.  We thus repeat our re-simulation process once more, using
the central $\sim10-30\,$pc of the first re-simulations to initialize
a new set of even smaller-scale simulations.  These typically have
$\sim10^{6}$ particles,
a spatial resolution of $0.1\,$pc, and a particle mass
$\approx100\,\msun$. We carried out $\sim50$ such simulations to test
the robustness of our conclusions and survey the parameter space of
galaxy properties.  These final re-simulations are evolved for
$\sim10^{7}$ years. 
We also carry out a few extremely
  high-resolution intermediate-scale simulations, which include
  $\sim5\times10^{7}$ particles and resolve structure from $\sim$ kpc
  to $\sim0.3\,$pc, obviating the need for a
  second zoom-in iteration.

Because of the one-way ``re-simulation'' method adopted here, 
the small-scale simulations are effectively instantaneous relative to the 
larger scale simulations. They capture the gas dynamics for one particular 
realization of the gas distribution on small and large scales, but cannot be 
evolved for long timescales relative to the large-scale simulations.


Our simulations include gas cooling and star formation, with gas
forming stars at a rate motivated by the observed \citet{kennicutt98}
relation ($\dot{\rho}_{\ast}\propto\rho^{1.5}$). 
%
%
Because we cannot resolve the detailed processes of supernovae
explosions, stellar winds, and radiative feedback, feedback from stars
is modeled with an effective equation of state 
\citep{springel:multiphase}. In this model, feedback is assumed to
generate a non-thermal (turbulent, in reality) sound speed; 
we use sub-grid sound speeds $\sim20-100\,{\rm km\,s^{-1}}$, motivated
by a variety of observations of dense, star forming galaxies \citep{downes.solomon:ulirgs,
  bryant.scoville:ulirgs.co,forsterschreiber:z2.disk.turbulence,
  iono:ngc6240.nuclear.gas.huge.turbulence}.
Within this range, we found little difference in the physics of
angular momentum transport or in the resulting BHARs, gas
masses, etc. \citep{hopkins:zoom.sims} (see also \S \ref{sec:results}). 
Critically, our model does {\em not} include any explicit ``fast stellar winds'' 
or time-dependence in the stellar feedback or mass recycling; there is an effective ISM pressure 
which corresponds to turbulent ISM sound speeds $<100\,{\rm km\,s^{-1}}$, 
but this is much less than the $\sim1000-2000\,{\rm km\,s^{-1}}$ O-star wind velocities 
and, more importantly, is time-independent. Therefore any characteristic timescale 
that emerges in these simulations is not a consequence of stellar evolution or 
mass recycling (as we have implemented it), even if this is 
in reality critical to explaining certain properties of the ISM 
\citep[see e.g.][]{hopkins:rad.pressure.sf.fb,hopkins:fb.ism.prop,hopkins:stellar.fb.winds}.

For all simulations, the total SFR within a given radius is 
directly determined by integration of the SFR in each gas particle 
\citep[this is well-converged in each case with respect to resolution; see][]{springel:models}. 
The BHAR requires more care. In the nuclear-scale simulations, 
the BHAR (which we assume is proportional to the AGN luminosity) 
is taken to be the inflow rate into a radius $<0.1\,$pc. 
This radius approximately represents the radius where we expect 
the traditional viscous $\alpha$-disk to begin \citep{goodman:qso.disk.selfgrav}, 
so the further inflow rate should be relatively constant at smaller radii 
\citep[see the discussion in][]{hopkins:zoom.sims}. 
In the intermediate and galaxy-scale simulations, we do not resolve the 
$<0.1\,$pc radii from which we would like to determine the BHAR 
(except in our few extremely high-res experiments). 
We therefore follow \citet{hopkins:zoom.sims} and ``map'' between scales, 
by taking the smaller-scale simulation most closely matched to the inner 
conditions in our large-scale simulation at each time (within some tolerance) 
and assuming this is a crude proxy for the average behavior on small scales 
given the large-scale conditions
(see \citealt{hopkins:zoom.sims} \S~6 for details). We do not intend to 
represent this as a literal ``zoom-in'' or exactly equivalent to using a 
single very high-res simulation. But using only our ultra-high resolution simulations 
(which obviate the need for this mapping) does not change our key 
qualitative conclusions (but is much noisier 
and covers a much smaller dynamic range, since there are only a few such 
simulations).

\vspace{-0.7cm}
\section{Results}
\label{sec:results}


\begin{figure}
    \centering
    \scaleup
    \plotone{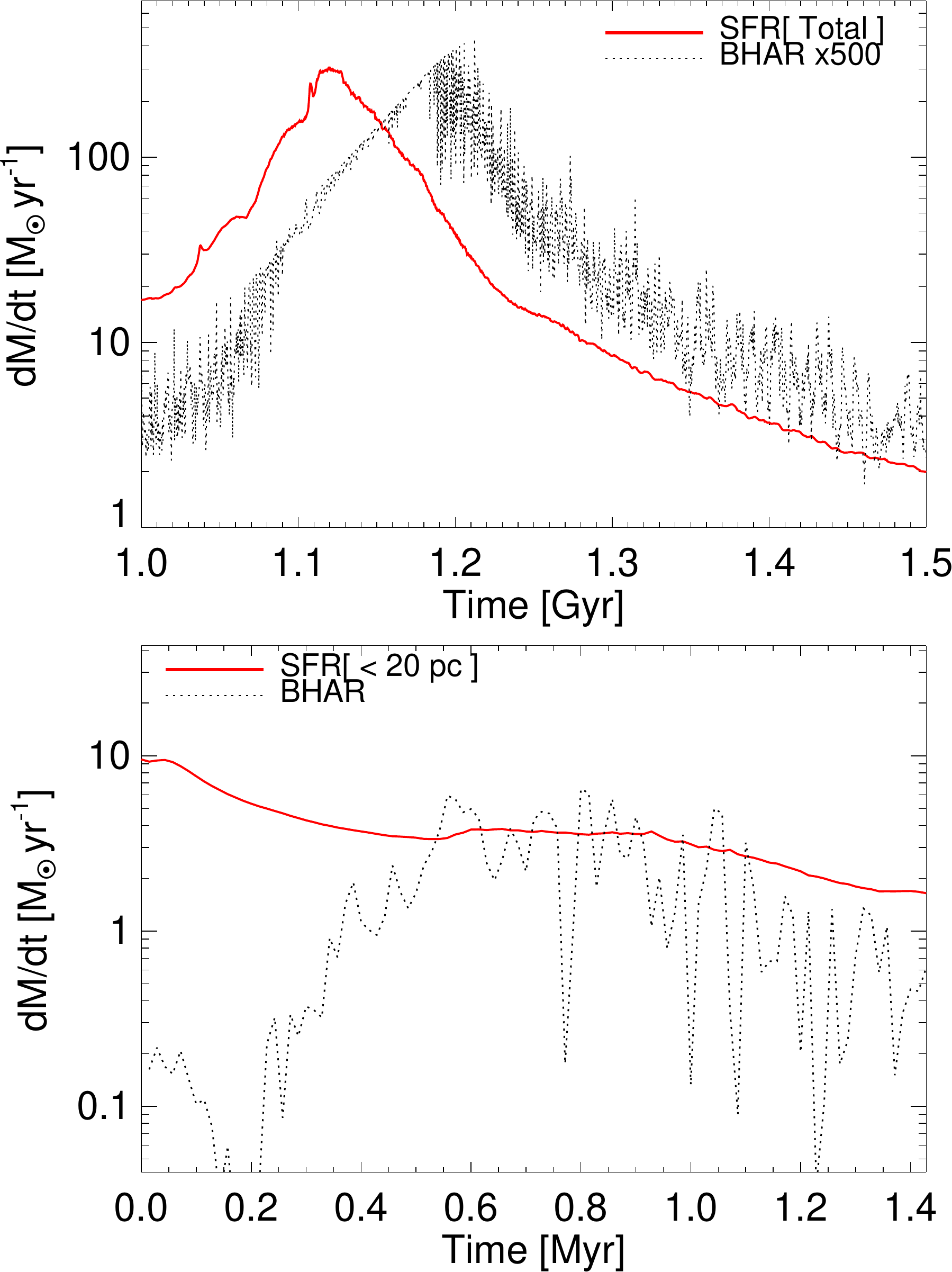}
    \caption{{\em Top:} Galaxy merger simulation, near coalescence of the two 
    galaxy nuclei (at $t\approx1.1\,$Gyr). The total SFR peaks near this 
    coalescence, as inflows first reach $\sim$kpc scales. The BH accretion 
    rate (here multiplied by 500 for ease of comparison) grows rapidly during this 
    time, but sufficient gas remains to fuel BH growth until $\sim10^{8}$\,yr 
    later. 
    {\em Bottom:} Zoom-in simulation of the central $<10\,$pc region of an 
    active system near the peak in BH accretion. The SFR on small scales 
    ($<20\,$pc) again peaks near the beginning of the simulation, when the 
    gas fraction in this radius is nearly unity and the gas has just reached small 
    radii. Inefficient angular momentum transfer in the gas, and the continued large 
    supply until most is exhausted, means that the AGN accretion rate 
    does not peak until $\sim$Myr later. 
    \label{fig:delay.ill}}
\end{figure}

Figure~\ref{fig:delay.ill} shows an example of two such simulations, with the 
BHAR and SFR evaluated on different scales. 
First, a typical gas-rich major merger simulation ($\fgas\approx0.4$ at time of 
merger, between equal-mass otherwise Milky-Way like disks).
We isolate the time near the 
final coalescence of the two galaxy nuclei, and plot the total SFR within the 
galaxy and the estimated BHAR. 
There is a 
clear offset. The merger drives strong gas inflows into the central kpc which 
drives the SFR to a peak at just about the time of coalescence. During this time, 
BH growth is accelerating rapidly, but there is more than enough gas to continue 
rapid BH growth even after the SFR turns over (i.e.\ the gas begins to be exhausted). 
The BHAR finally turns over about $10^{8}\,$yr later. 
Second, we show one of our nuclear-scale zoom-in simulations,  
appropriate for the central regions of this simulation near the peak in the 
BH activity (initial $\fgas=0.8$ at small radii 
with a $3\times10^{7}\,\msun$ BH: Nf8h1c1dens
in \citealt{hopkins:zoom.sims}). 
We compare the SFR within some small resolved radius ($<20\,$pc here), 
where the inflows have just reached (hence the large initial gas fraction), 
to the BHAR (gas inflow rate into the central $<0.1\,$pc region). 
The SFR peaks immediately (more appropriately it would have peaked even 
slightly before the beginning of this simulation), as it simply traces the gas supply. 
But the inflows to small radii take some time to become maximal, $\approx1\,$Myr 
here. 

\begin{figure}
    \centering
    \scaleup
    \plotone{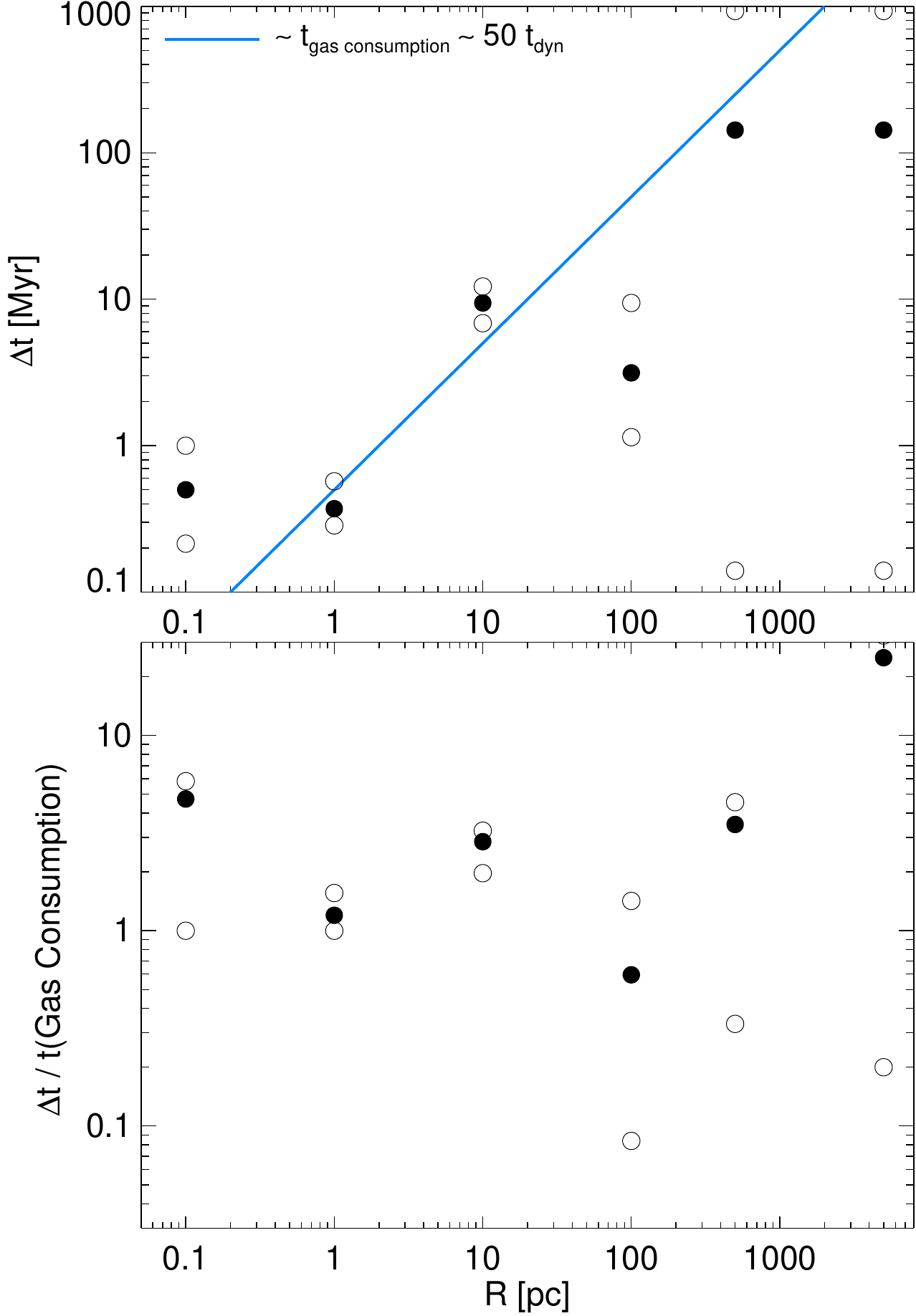}
    \caption{{\em Top:} Delay time $\Delta t$, between the peak of gas inflow 
    and star formation into a given scale ($<R$), and the peak of 
    AGN activity (inferred inflow rate into $<0.1\,$pc), measured in a suite 
    of hydrodynamic simulations with resolution ranging from 
    $0.1\,$pc (leftmost two points) to $3$\,pc (middle two) to $50\,$pc (right two). 
    The blue line compares a timescale crudely similar to $\epsilon^{-1}\,t_{\rm dyn}$ 
    where $\epsilon\approx1/50$ (the SF efficiency per dynamical time) and 
    $t_{\rm dyn}\approx R/{100\,{\rm km\,s^{-1}}}$ are assumed. 
    Filled circles show the median $\Delta t$ in a sample of simulations, 
    open circles correspond to the scatter (the $25-75\%$ interval). 
    {\em Bottom:} Same, but for the ratio $\Delta t/t({\rm Gas\ Consumption})$, 
    where $t({\rm Gas\ Consumption})$ is defined as the 
    time between when the gas fraction inside $R$ peaks (usually here 
    at $\fgas=0.8-1$) from inflows from larger scales or initial conditions, 
    and when it falls below a value $\fgas\le 0.2$. 
    \label{fig:delay.summary}}
\end{figure}

Figure~\ref{fig:delay.summary} extends this to our ensemble of simulations, 
plotting the delay $\Delta t$ between the peak of AGN activity and the 
peak of star formation activity within a given radius $R$. 
The points at $0.1-10$\,pc use just each nuclear-scale simulation individually; 
those at $100-500$\,pc use the intermediate-scale simulations 
(including ultra-high resolution runs on this scale which do not need another 
``zoom in'' iteration to resolve the BHAR); 
those at $>500\,$pc use the galaxy-scale simulations. 
In each case we measure the total SFR within the annulus plotted, 
and compare it as a function of time to the estimated BHAR, each smoothed 
over a timescale corresponding to a dynamical time at the radius $R$ 
(since both are strongly time-variable).\footnote{The points which correspond to 
the outermost scales in each tier of simulations should be regarded with some 
caution, as the characteristic timescales shown could be influenced by how 
the simulations were re-scaled (since the outer boundary conditions 
are not necessarily identical to a single, arbitrarily high-resolution simulation). 
However, the inner regions can be reliably evolved for many dynamical times. 
We have also compared the values plotted with the results from 
our more limited suite of extremely high-resolution intermediate-scale simulations 
that ``bridge'' multiple scales and obviate the need for rescalings; these 
are consistent in all cases.}

There is a clear delay which is larger 
when the SFR is measured at larger radii; this arises for two reasons.
First, angular momentum loss in gas from gravitational instabilities 
occurs via transfer to the stellar material, when 
the magnitude of asymmetries in the potential is sufficient to induce shocks and 
dissipation in the gas \citep{hopkins:disk.survival,hopkins:zoom.sims,
hopkins:inflow.analytics}. 
In the limit of a pure gas system, there is no collisionless 
component to absorb the angular momentum, and so the leading-order angular 
momentum loss is reduced to second-order resonance effects \citep{kalnajs:1971} 
or an effective (turbulent) viscosity \citep{gammie:2001.cooling.in.keplerian.disks}. 
To leading order, the efficiency of 
angular momentum exchange in some annulus scales as 
$\propto (1-\fgas)$, where $\fgas = M_{\rm gas}/(M_{\rm gas}+M_{\ast})$ is 
the local gas fraction in the disk \citep{hopkins:inflow.analytics}.\footnote{Spherical distributions of 
material such as the bulge and halo enter in separable fashion.}
Inflows that ``pile up'' high $\fgas$ in $\sim t_{\rm dyn}$ in simulations therefore tend to 
stall until $\fgas$ can be lowered (the longer gas exhaustion timescale). 


Second, it requires much less gas to fuel an AGN, as opposed to a starburst. 
The SFR inside an annulus $R$ 
peaks when the gas mass inside that radius is maximum. In contrast, the 
total mass needed (in principle) to fuel the BH is very small. Even if ``efficiencies'' 
of gas getting to the BH are low, once sufficient fuel exists to grow the BH at the 
Eddington limit, having more gas at large radii will not further increase the BHAR. 
For a typical starburst, the gas mass within $\sim500\,$pc at the time of peak SFR is 
$\gtrsim100$ times that needed to fuel the BH. It is therefore at least possible that 
the BH could continue to grow at or near its Eddington limit (so the 
luminosity will continue to increase exponentially) until $>99\%$ of the 
gas is exhausted, which for typical efficiencies could require $\gg100$ dynamical 
times. Of course, accretion is not so efficient.
But in previous simulations of galaxy mergers with simple prescriptions for 
BH accretion proportional to either the Bondi rate \citep{springel:models}, 
viscous accretion rate \citep{debuhr:momentum.feedback}
or gravitational instability accretion rate \citep{hopkins:inflow.analytics} estimated on 
$\sim100\,$pc scales, the gas supply on this scale in the late merger stages is so large that the estimated 
BHAR is at or above Eddington throughout the entire nuclear starburst 
(i.e.\ luminosity rises rapidly), until $>10^{8}$\,yr after 
the SFR peaks and begins to decline from gas exhaustion. 
\citet{hopkins:inflow.analytics} argue that in the case of inflows in gas+stellar disk, 
the BHAR scales (on average) $\propto 1/(1+f_{0}/f_{\rm gas})$
where $\fgas$ is measured on large scales and $f_{0}\sim0.1$ depends on the 
details of the system -- the important point is that this implies accretion 
relatively independent of $\fgas$ in the large-$\fgas$ limit, but then 
cuts off when $\fgas\ll 1$. 

Both these arguments imply that the 
critical timescale that determines the delay between 
peak SFR and AGN activity is the gas exhaustion time. 
The Kennicutt-Schmidt relation implies an exhaustion time $\Delta t(R)$
\be
\Delta t(R) = \frac{\Sigma_{\rm gas}}{\dot{\Sigma}_{\ast}}
\approx
\frac{\Sigma_{\rm gas}}{\epsilon\,\Sigma_{\rm gas}\,t_{\rm dyn}^{-1}}
=\epsilon^{-1}\,t_{\rm dyn}
\ee
where $\dot{\Sigma}_{\ast}$, $\Sigma_{\rm gas}$, $t_{\rm dyn} =1/\Omega= R/V_{c}$ are the SFR 
density, gas surface density, and orbital frequency, and 
$\epsilon$ is the efficiency (fraction of gas turned into stars per dynamical time). 
Observations suggest values of $\epsilon\sim0.006-0.1$, approximately 
independent of local density \citep{krumholz:sf.eff.in.clouds,
bigiel:2008.mol.kennicutt.on.sub.kpc.scales,leroy:2008.sfe.vs.gal.prop}.


We test this scaling in Figure~\ref{fig:delay.summary}.
We compare the measured $\Delta t(R)$ with a crude estimator of the 
dynamical time, $\sim 0.5\,{\rm Myr}\,(R/{\rm pc})$.
This is the expectation from our above derivation if 
the system had a flat rotation curve with $V_{c}=100\,{\rm km\,s^{-1}}$, 
and $\epsilon\sim0.02$. 
Specific values will depend on the galaxy properties as well as the 
adopted star formation law (related to the uncertain sub-grid SF law). 
More directly, we therefore compare with the gas consumption time 
$t(\rm Gas\ Consumption)$, where 
$t(\rm Gas\ Consumption)$ is defined as the time 
from when $\fgas$ peaks within a given annulus (about the peak SFR time) until it 
falls to $f_{\rm gas}<0.2$ (the exact value is arbitrary 
so long as $\fgas\ll1$, 
we choose this because it gives about the right normalization). 
We measure $\Delta t/t(\rm Gas\ Consumption)$ for each simulation, 
and plot the distribution of this ratio as a function of $R$. 
There is (unsurprisingly) large scatter at larger radii especially, but this 
provides a decent approximation to the simulation results. 


\vspace{-0.7cm}
\section{Discussion}
\label{sec:discuss}

Simulations of AGN fueling by gravitational instabilities naturally produce a 
delay between the time when 
SFRs peak inside of a given annulus and the time when 
AGN activity peaks. This offset scales as the 
gas consumption time, $\sim10-100$ 
dynamical times. On small scales ($\lesssim10\,$pc), this is characteristically 
$\sim10^{7}\,$yr, rising to a few $\sim10^{8}$\,yr on kpc scales. 
These offsets are similar to the magnitude of time offsets suggested 
by various observations on both small and large 
scales \citep{davies:sfr.properties.in.torus,
schawinski:agn.sf.suppression.timescale,wild:agn.sf.offset}.

Of course, both the AGN luminosity and SFR are 
highly time-variable \citep[see e.g.][]{levine:sim.mdot.pwrspectrum}. 
Therefore this should be taken as a statement only about time-averaged 
accretion rates and SFRs on a timescale of at least a few dynamical times. 
There will be considerable variation from one system to 
another. 
And even in a time-averaged sense, the simulations show a range of behaviors, 
with the magnitude of apparent ``time offsets'' varying from $\sim 0.1-10$ times 
the gas consumption time (on large scales, from $\sim$Myr to Gyr timescales); 
there are even some systems where the AGN luminosity peaks before the SFR. 

Nevertheless, this may be sufficient to account for current observations 
without invoking poorly-understood stellar fueling (or feedback) processes. 
Those processes may still be important, of course, and in future 
work we will investigate how the explicit stellar evolution models presented 
in \citet{hopkins:rad.pressure.sf.fb,hopkins:fb.ism.prop,hopkins:stellar.fb.winds} 
alter the dynamics discussed here. Observationally 
testable consequences of these models include the magnitude of 
time offsets, their scaling with radii and the dynamical time of the 
galaxy, and the scaling of AGN activity with gas fraction inside some radius. 
Since it is gas fraction that drives these effects, one does not expect to 
see evidence for AGN -- on average -- reaching their peak luminosity 
while the gas fraction on large scales is very large $\fgas\approx1$, 
but rather expects to see that the ``delayed'' peak in AGN activity 
corresponds to systems with $\fgas\lesssim0.1-0.5$. 
And there is some observational evidence from 
metal enrichment and stellar age profiles that high gas densities and 
star formation proceed ``outside-in'' in similar fashion to that predicted here 
in the early stages of mergers \citep{soto:ssp.grad.in.ulirgs,
kewley:2010.gal.pair.metal.grad.evol,rupke:2010.metallicity.grad.merger.vs.obs,
torrey:2011.metallicity.evol.merger}.

This also has important implications for AGN feedback models. 
AGN luminosities in realistic hydrodynamic models do {not} 
peak at the same time as the gas density on small scales, 
but rather at somewhat later times when {most} of the gas has been exhausted 
by star formation and/or expelled by stellar feedback. 
To the extent that AGN feedback assists in the 
``blowout'' of gas (shutting down star formation), it primarily 
sweeps up the ``trickle'' of remaining gas that would provide 
for low-level star formation over the next few Gyr, 
not the bulk of the galaxy gas supply \citep[see e.g.][]{hopkins:groups.qso,
hopkins:groups.ell}. 
For the simulations shown here, it has been shown previously 
that the AGN feedback actually acts on a very 
small residual amount of gas, and has a nearly negligible effect 
on the peak-SFR nuclear starburst \citep[see][]{cox:kinematics,robertson:fp,
hopkins:cusps.fp}. 
Similar conclusions have recently been obtained  
using entirely different accretion and AGN feedback models 
\citep{debuhr:momentum.feedback}. 
But this may be needed to explain the rapid color evolution 
of galaxies through the ``green valley'' \citep[see e.g.][]{kaviraj:2011.agn.fb.needed.in.postsb}.

This also means that galaxies at their peak in star formation 
activity may not necessarily be ideal places to look for the 
effects of AGN feedback. Indeed, observations of such systems have 
found mixed results \citep[see e.g.][and references therein]{
veilleux:winds,rupke:outflows,krug:agn.feedback.vel.vs.lir}. 
Post-starburst \citep{tremonti:postsb.outflows,
wild:postsb.fb.evidence.and.gives.all.ell,coil:2011.postsb.winds} or 
later-stage, AGN-dominated mergers \citep[]{sturm:2011.ulirg.herschel.outflows,
humphrey:2010.type2.qso.feedback,dunn:agn.fb.from.strong.outflows,
bautista:2010.strong.agn.fb,feruglio:2010.mrk231.agn.fb} 
might be more promising.

We thank Vivienne Wild, Kevin Schawinski, Ric Davies, and Tim Heckman for 
helpful discussions that inspired this manuscript. We also thank the anonymous 
referee for a number of thoughtful suggestions.

\vspace{-0.23in}
\bibliography{/Users/phopkins/Documents/lars_galaxies/papers/ms}

\end{document}